# Assessing scientific research performance and impact with single indices

J. Panaretos[1] and C.C. Malesios
*Department of Statistics*
*Athens University of Economics and Business*
*76 Patision St, 10434 Athens Greece*

**Abstract**

We provide a comprehensive and critical review of the h-index and its most important modifications proposed in the literature, as well as of other similar indicators measuring research output and impact. Extensions of some of these indices are presented and illustrated.

## 1. Introduction

The need for accountability in Higher Education (HE) has led governments, research authorities and University administrators to assess research performance using single indices that allow comparisons and rankings. Characteristically, the UK government has recently decided to replace, after 2008, the current method for determining quality in HE (the research assessment exercise). Metrics[2], rather than peer review will be the focus of the new system and it is expected that bibliometrics (using counts of journal articles and their citations) will be the central quality index of the system [see *Evidence Report, 2007*].

Rankings of Higher Education Institutions based on such single indices appear frequently in the media generating concerns in Institutions and national Governments. Even an EU-commissioned report [*Saisana and d'Hombres, 2008*], although accepts the inevitability of such rankings, it argues that popular world rankings such as the Academic Ranking of World Universities published by the Shanghai Jiao Tang University (SJTU) and the THES-QS World University Rankings published by the

---

[1] e-mail for correspondence: jpan@aueb.gr
[2] The UK government has indicated that mathematics and statistics will not be included in the first phase of the shift to metrics

Times Higher Education supplement and Quacquarelli Symonds (THES), are highly sensitive to methodological assumptions.

The concern for the implications of poor performance in such rankings has led Governments to consider taking some action. As mentioned in the report quoted above, the French President Sarkozy has stressed the need for French Universities to consolidate in order to promote their ranking. Also, because of the political importance of higher education rankings, the French ministry of Education is considering the creation of a new University ranking system.

Institutions on the other hand, are making efforts to improve their standing in the rankings. For example, the Swiss Federal Institutes of Technology at Zurich and Lausanne has issued instruction to their faculty members on using a uniform way to state their affiliation in their publications so that no paper is "lost" by misattribution. (At least three different ways of stating the affiliation for each of them has been observed in the literature).

All of the above inevitably led to criticisms of the use of such "simple" measures of research performance. A recent report by the joint Committee on Quantitative Assessment of Research [*Adler et al., 2008*] argues that "research is too important to measure its value with only a single coarse tool".

It is natural therefore that research associated with the assessment of the advantages, disadvantages and limitation of such indices is growing.

Up to 2005, the traditional bibliometric indicators were based on simple statistical functions, for instance, means, relative frequencies and quantiles [*Glänzel, 2006*]. One of the main disadvantages of standard bibliometric indicators, such as the total number of papers or the total number of citations, is that they do not reflect the full impact of scientific research, or that they are disproportionately affected by a single publication of major influence.

In 2005, a new indicator for the assessment of the research performance of scientists was proposed by *Hirsch [2005]*, intended to measure simultaneously the quality and sustainability of scientific output, as well as, to some extent, the diversity of scientific research. The specific index attracted interest immediately and has received a lot of attention [see, e.g. *Ball, 2005*]. Since its introduction, a long series of articles has appeared, proposing modifications of the original h-index for its improvement, or implementations.

The h-index *[Hirsch, 2005]* is an index built to consider both the actual scientific productivity and the scientific impact of a scientist. The index is based on the set of the scientist's most quoted papers and the number of citations that they have received in other scientists' publications. The specific index has also been applied to the productivity and impact of a group of scientists, such as a department, a university or a country. The index was suggested as a tool for determining the relative quality of research by theoretical physicists, and is sometimes called the Hirsch index (also met as the Hirsch number in the literature).

Hirsch has argued that h has a high predictive value as to whether a scientist has won honors like the National Academy membership or the Nobel Prize. He has also calculated the h index from the 10 most highly cited researchers from the field of biomedical sciences, and found that all highly cited researchers also have high h-index numbers. Large differences appear among different scientific disciplines, as concerns the magnitude of the h-index (see section 4 for more details).

**Definition:**

*A scientist has index h if h of his $N_p$ papers have at least h citations each, and the other ($N_p$ - h) papers have at most h citations each.*

Thus, for example, a scientist with an index of 10 has published 10 papers with at least 10 citations each. A zero h-index characterizes authors that have at best published papers that have had no visible impact *[Glanzel, 2006]*. The papers that contribute to the calculation of the h-index (i.e. the papers receiving h or more citations) are referred to as the h-core papers.

A mathematical-based definition [see, e.g. *Glanzel, 2006*] can be given as follows: Consider an author who has published a series of n papers, where the ith paper (i=1,2,…,n) has received $X_i$ citations. If we order the number of citations of the n articles in a decreasing order, we have:

$$X_1^* \geq X_2^* \geq \ldots \geq X_n^*,$$

where $X_1^*$ denotes the number of citations received by the most cited paper and $X_n^*$ denotes the number of citations received by the less cited paper. Under this setting:

$$h = \max\{j : X_j^* \geq j\}.$$

Thus, the h-index is the result of the balance between the number of publications and the number of citations per publication. The index is designed to improve upon simpler measures such as the total number of citations or publications, to distinguish truly influential (in terms of citations) scientists from those who simply publish many papers. Among the advantages of this index is its simplicity, the fact that it encourages researchers to produce high quality work, that it can combine citation impact with publication activity and that is also not affected by single papers that have many citations. Another attractive property of the h-index is that it is robust to small cited publications, i.e. just an increase in the number of publications does not improve the h-index.

However, despite the potential of this index, one may argue that more work should be done on both the theoretical aspect and the applications of this index.

## 2. Bibliographic Data Sources

Online web programs are available to directly calculate a scientist's h-index, for instance QuadSearch, a Metasearch Engine provides an h-index calculation and related charts (http://quadsearch.csd.auth.gr/). Alternatively, there are free internet citation databases from which h can be manually determined, such as Google Scholar (http://scholar.google.gr/). Subscription-based databases such as Scopus (http://www.scopus.com/scopus/home.url) and the Web of Knowledge (http://www.isiwebofknowledge.com/) provide automatic functions and more complete databases. Each of the above databases, however, is likely to produce a different h-index for the same academic scientist. This has been studied in various articles (see, e.g., *Kosmulski [2006], Bornmann and Daniel [2007], Jin et al. [2007]*).

According to other authors, the Web of Knowledge was found to have strong coverage of journal publications, but poor coverage of high impact conferences (a particular problem for Computer Science based scholars); Scopus has better coverage of conferences, but poor coverage of publications prior to 1992; Google Scholar has the best coverage of conferences and most journals (though not all), but like Scopus has limited coverage of pre-1990 publications. As *Meho [2007]* reports from the results of a study in the field of information science, Google Scholar and Scopus can increase citation counts by an average of 160 per cent and 35 per cent respectively, compared to the Web of Knoweledge. Google Scholar has also been criticized for including gray literature in its citation counts, since in addition to published papers it includes citations to working papers and books, among other sources. However, a study showed that the majority of the additional citation sources of Google Scholar are contributed by legitimate refereed forums [*Meho and Yang, 2007*].

Another problem associated with the use of such databases for calculating the h-index is that of discriminating scientists that share the same first and/or last names. When searching for papers by a scientist by means of only the author search field in the Web of Science database for instance, it cannot be ruled out with certainty that papers by a different scientist with the same last name are not counted into the calculation of the h-index. To overcome this deficiency, one needs to manually calculate the h-index, to exclude, for instance, citations received by scientists with the same last name. This is time-consuming.

It should be stressed that the content of all of the databases, particularly the Google Scholar, continually changes, so any research on the content of the databases risks going out of date.

A disadvantage of the Google Scholar presented in the literature [see, e.g., *Sanderson, 2008; Meho, 2006*] is that with Google Scholar, the number of citations is inflated, when compared to other citation sources. Nevertheless, it is suggested that in order to deal with the sometimes wide variation in h-index for a single academic measured across the possible citation databases, it is better to take into account the maximum h-index measured for a researcher rather than the minimum *[Sanderson, 2008]*. *Meho [2007]* suggests the use of more than one citation sources in order to make correct comparisons and derive accurate assessments of the publication output of a researcher.

More recently, a specialized software program ("Publish or Perish") *[Publish or Perish User's Manual, 2007]* has been released, which collects and analyzes citation data using Google Scholar. In addition to the numbers of articles and citations of researchers, the software calculates a series of Hirsch type indicators, such as the h-index, Egghe's g-index, as well as other variations of the index proposed by *Hirsch [2005]*. An automatic calculation of h-index is also recently provided by the Web of Science, using the "citation report" function, reducing thus the time required for its manual calculation.

Our opinion (based on our experience of the use of various citation-collecting sources) is similar to that of *Meho [2007]* – that the use of more than a single source of citation data in calculating h-indices is a necessary (still not sufficient) condition to derive valid results. Combination of web citation sources and databases with other sources of information (for instance the scientists' personal web pages or University

Departments' pages when available) can help in the effort to clean raw data containing irrelavent publications/citations or duplicate records.

Exclusion of irrelavent citations can be also achieved by including in the search field the author's two initials of his/hers first name. However, caution is required in doing this, since valid citations could be left uncounted because, in many occasions, the author's name appearing in an article includes only the first initial. Other proposals including checking the researcher's affiliations [*Schreiber, 2007*], are not easily applicable, since movement of researchers between Institutions/Univesrities is a very frequent practice.

**3. Disadvantages of the h-index**

The popularity and the wide use of the h-index has raised a lot of criticism, also. As we have already mentioned, the most notable and well-documented example of critical view on the h-index (and other "simple" measures of research performance[3]) is a recent report by the joint Committee on Quantitative Assessment of Research [*Adler et al., 2008*]. In this report, the authors argue strongly against the use (or misuse) of citation metrics (e.g., the impact factor or the h-index) alone as a tool for assessing quality of research, and encourage the use of more complex methods for judging scientists, journals or disciplines, that combine both citation metrics as well as other criteria such as memberships on editorial boards, awards, invitations or peer reviews (the interested reader can also refer to *Moed and van Leeuwen [1996]* and *MacRoberts and MacRoberts [1989]* for a thorough discussion on the criticism of impact factors, and citation metrics in general).

With regard to the h-index (and associated modifications), specifically, *Adler et al. [2008]* stress that its simplicity is a reason for failing to capture the complicated citation records of researchers, loosing thus crucial information essential for the assessment of a scientist's research. The lack of mathematical/statistical analysis on the properties and behaviour of the h-index is also mentioned. This is in contrast to the rather remarkable focus of many articles to demonstrate correlations of h-index with other publication/citation metrics (i.e. published papers or citations received), a result which according to the authors is self-evident, since all these variables are essentially functions of the same basic phenomenon, i.e. publications.

---

[3] The report also outlines inefficiencies of citation metrics such as the Impact Factor (IF), especially when implemented to assess academic quality of journals in the field of mathematics.

The criticism by *Adler et al.* is not solely targeted towards the h-index, but includes all relevant metrics that use citation data in their calculation. Following *Cozzens [1989]*, who argues that citations are the result of two systems, one of which is the "reward" system and the other is the "rhetorical" system, the authors point out the complexity of citations, stating that a citation cannot be counted a priori as an acknowledgment of a scientist's work, since there can be many other reasons that can create a citation, such as the negative (or "warning") citation, or a citation that explains some result, or even a self-citation. For example, citations in a paper are often made simply to flesh-out an introduction, having no other reference to the essence of the work. As already mentioned, this disadvantage also characterizes other metrics that use citations (the interested reader is referred to *Martin and Irvine [1983]* for more on this subject). An in-depth discussion of the prominent issue of performance indicators in general and the appropriateness of their use in comparing various sectors of human activity, such as education, health system or social services can be found in *Goldstein and Spiegelhalter [1996]*.

Besides the above-mentioned works, there are many more articles referring to disadvantages of the h-index. In what follows we list some of the most important disadvantages of the h-index:

- The h-index is bounded by the total number of publications. This means that scientists with a short career (or at the beginning of their career), are at an inherent disadvantage, regardless of the importance of their discoveries. In other words, it puts newcomers at a disadvantage since both publication output and citation rates will be relatively low for them (see, e.g. *Sidiropoulos et al. [2006]*).

- Some authors have also argued that the h-index is influenced by self-citations *[Hirsch, 2005; Schreiber 2007a; Vinkler, 2007]*. Many self-citations would give a false impression that the scientists' work is widely accepted by the scientific community. According to *Vinkler [2007]*, both self-citations and "real" (independent) citations are usually used in the calculation of the h-index. In this context, the emerging problem is that scientists with many co-operating partners may receive many self-citations, in contrast to scientists that publish alone. (*Meho [2007]* refers to the problem of exchanging citations between collaborating scientists using the term "cronyism").

- The h-index has slightly less predictive accuracy and precision than the simpler measure of mean citations per paper [*Lehmann et al., 2006*].

- Another problem is that the h-index puts small but highly-cited scientific outputs at a disadvantage. While the h-index de-emphasizes singular successful publications in favor of sustained productivity, it may do so too strongly. Two scientists may have the same h-index, say, h = 30, i.e., they both have 30 articles with at least 30 citations each. However, one may have 20 of these papers that have been cited more than 1000 times and the other may have all of his/hers h-core papers receiving just above 30 citations each. It is evident that the scientific work of the former scientist is more influential. Several recipes to correct for this have been proposed, but none has gained universal support [see, e.g., *Egghe [2006a, 2006b, 2006c], Kosmulski [2007]*].

- Limitations/differences of the citation data bases may also affect the h-index. Some automated searching processes find citations to papers going back many years, while others find only recent papers or citations (see *Sanderson [2008]* for a detailed data base comparison).

- Another database related problem often occuring with a significant effect on the correct calculation of the h-index, is that of name similarities between researchers. (*Meho [2007]* uses the term "homograph" to describe failure to separate scientists sharing the same last name and initials). As *Kosmulski [2006]* stresses out, it is almost impossible to find a scientist with a unique combination of family name and initials while searching the most known citation databases. As a result, in many cases the h-index will be overestimated, since in its calculation the works of more than one researcher are added[4]. In a recent article, *Jacsó [2008a]* [see also *Jacsó, 2008b*], using as an example the name of a distinguised scientist from the field of Information science, compares extensively the most commonly used citation sources Goggle Scholar, WoS and Scopus and concludes that all databases suffer from significant insufficiencies, mainly in the accuracy of the calculation of the h-index.

- It seems that the h-index cannot be utilized for comparing scientists working in different scientific fields. It has been observed that average citation numbers differ widely among different fields [*Hirsch, 2005; Podlubny, 2005*].

---

[4] According to Hirsch [2007], the Web of Science database has recently alleviated this problem by incorporating specialized discriminating tools under the "author finder" option of the database.

- General problems associated with any bibliometric index, namely the necessity to measure scientific impact by a single number, apply here as well. While the h-index is one 'measure' of scientific productivity, some object to the practice of taking a human activity as complex as the formal acquisition of knowledge and condense it to a single number *[Kelly and Jennions, 2006]*. Two potential dangers of this have been noted: (a) Career progression and other aspects of a human's life may be damaged by the use of a simple metric in a decision-making process by someone who has neither the time nor the intelligence to consider more appropriate decision metrics. (b) Scientists may respond to this by maximising their h-index to the detriment of doing more quality work. This effect of using simple metrics for making management decisions has often been found to be an unintended consequence of metric-based decision taking; for instance, governments routinely operate policies designed to minimize crime figures and not crime itself.

**4. Some Generalizations/ Modifications of the h-index**

Soon after the introduction of the h-index, various modifications and generalizations of it have appeared in the literature. Most of them are indented to correct the insufficiencies of the h-index, already described in the previous section. Among them, we can distinguish the g-index [*Egghe, 2006a, 2006b, 2006c*], the R- and AR-index [*Jin et al., 2007*], the A-index [*Jin, 2006*], the contemporary, trend and normalized h-index [*Sidiropoulos et al., 2006*] and the $A^{(2)}$ index [*Kosmulski, 2006*].

In the following, we attempt to review the recent literature on the work associated with modifications of the h-index.

*4.1 h-type indices adjusting for the robustness of h-index to the number of h-core citations*

As already stressed, the h-index has been reported [see, e.g., *Egghe, 2006*] as being totally robust to variations of the number of citations received by the papers belonging to the h-core. In order to account for this "robustness", various modifications appeared in the literature. These include the g-index [*Egghe, 2006a, 2006b, 2006c*], the A-index [*Jin, 2006*], the R-index [*Jin et al., 2007*], the $h_w$-index [*Egghe and Rousseau, 2007*], the w-index [*Wu, 2008*] and the $A^{(2)}$ index *[Kosmulski, 2006]*. In the following, these indices are presented, accompanied by a short example of their application using an artificial dataset. Some comments on the advantages/disadvantages of the indices are provided, along with some recommendations on the suitability of their use, depending on the occasion.

*The g-index*

The h-index is a robust index in the sense that it is insensitive to an accidental set of uncited (or lowly cited) papers and also to one, or several, outstandingly high cited papers. However, it is not sensitive to the level of the highly cited papers. Indeed, suppose a scientist has an h-index of value 10. Then for an article belonging to the h-core of this scientist, it is unimportant whether it has 10 or 100, or even 10000 citations. In order to overcome this, *Egghe [2006a, 2006b, 2006c]* defined the g-index[5].

Definition: *The g-index is the highest number g of articles that together received $g^2$ or more citations.*

Clearly, g≥h. By its definition, this index is increased by a strongly skewed frequency distribution of the citations, that is the higher the number of the citations in the top range, the higher the g-index.

*Egghe* [*2006b*] also presents two real author examples to illustrate the potential advantages of his proposed g-index.

The g-index clearly overcomes the h-index's insufficiency of not depicting the internal changes of the Hirsch core. Yet it requires drawing a longer list than necessary for calculating the h-index, hence it increases the precision problem. *Rousseau [2006a, 2006b]* studies the g-index, investigating its relation to the h-index using some simple models.

*The A-index*

*Jin's [2006]* A-index achieves the same goal as the g-index, namely correcting for the fact that the original h-index does not take into account the exact number of citations included in the h-core. It is simply defined as the average number of citations received by the articles included in the Hirsch core. i.e, $A = \frac{1}{h}\sum_{j=1}^{h} citation_j$.

In the above formula the numbers of citations are ranked in decreasing order. The A-index uses the same data as the h-index. It is obvious that always h≤A. The A-index suffers from the problem of punishing among scientists with the same A-index the one with higher h-index, since the A-index involves a division by h.

---

[5] Also introduced independently by Jin [2006]

*The R-index*

*Jin et al. [2007]* introduced and studied the R- index, which according to the authors, eliminate some of the disadvantages of the h-index. The R-index is an improvement of the A-index. Specifically, the R-index tries to eliminate the disadvantage of the A-index, by calculating the square root of the sum of the h citations included in the Hirsch core, i.e.: $R = \sqrt{\sum_{j=1}^{h} citation_j}$.

As one can observe, $R = \sqrt{Ah}$. It is also clear that h≤R.

*The $h_w$-index*

Another h-type index that aims at being sensitive to variations in the h-core is the $h_w$-index, defined by *Egghe and Rousseau [2007]*. The authors define the $h_w$-index in a discrete and a continuous setting, and establish a series of properties of the theoretical $h_w$-index in both settings.

To construct the $h_w$-index in the discrete case (the most practical of the two) one has to calculate the weighted ranks $r_w(j) = \sum_{i=1}^{h} C_j / h$, where $C_j$ denotes the number of citations received by the jth article, and h is the Hirsch index. Then, substitute with this weighted ranking the journal ranking according to the citations received used for the calculation of the h-index. The new index is given by $h_w = \sqrt{\sum_{i=1}^{r_0} C_i}$, where $r_0$ is the largest $r_w$-value such that $r_w(j) \leq C_j$. By applying this weighted ranking to the citations, *Egghe and Rousseau [2007]* introduced an index that takes into account the overall number of h-core citations as well as the distribution of the citations in the h-core.

*The $A^{(2)}$ index*

*Kosmulski [2006],* in an attempt to circumvent the problems of name similarities between researchers that reduces the precision in the calculation of h-index [see also *Jin et al., 2007*], introduced the $h^{(2)}$-index, defined as follows:

*A scientist has $h^{(2)}$-index, say k, if k of his Np papers have at least $k^2$ citation, and the other ($N_p$-h) papers have at least $h^2$ citations.*

Obviously, $h^{(2)} \leq h$ for any scientist. According to *Kosmulski [2006]*, $h^{(2)}$ is highly correlated with the total number of citations received by a scientist. *Liu and Rousseau [2007]* study the h- g- and $h^{(2)}$-indices and their use as indicators in a library management setting. They deduce – through practical implementations on a real dataset – that the $h^{(2)}$ index lacks in discriminatory power when utilized to assign ranks to different classes of books (playing the role of an author) and the loans on them (playing the role of citations received), when compared to the other two Hirsch-type indices.

*The w-index*

Another Hirsch-type index was recently proposed by *Wu [2008]*, and is called w-index. By definition: *A scientist has a w-index if w of his/hers papers have at least 10w citations each, and the remaining papers have fewer than 10(w+1) citations.*

As the author argues, the w-index appears to be very similar to the h-index, however it better reflects the influence of a scientist's top papers (for obvious reasons, w-index is alternatively called the 10h-index). *Wu [2008]*, examines the accuracy and properties of the w-index through an empirical analysis using bibliometric data on 20 astrophysicists. By practical implementations, the author found that it is $h \cong 4w$.

*The $i \times c_i$ index (maxprod index)*

In the context of h-type indices, the maxprod index is introduced by *Kosmulski [2007]*. Maxprod is defined as: *"the highest value among values $i \times c_i$, where i denotes the ith article and $c_i$ is the number of citations received by the ith article"*.

Maxprod is related to h-index, as follows: $maxprod \geq h \times c_h \geq h^2$. According to *Kosmulski [2007]*, the specific index has an advantage over the h-index (reported to be too robust to large differences in the number of citations in the h-core), since it can be utilized as a selective tool for identifying scientists of outstanding achievements (referred to as "genies" by the author) among the vast majority of scientists of "typical" scientific behavior. By using artificially constructed publication/citation distributions and real case studies, *Kosmulski [2007]* notices that for an outstanding scientist a typical $i \times c_i$ value is usually observed for i‹‹h, while for the typical scientist it is usually i≈h. Cases where i››h are characteristic of scientists that produce a lot articles, receiving only a few citations.

*The t- and f-indices*

*Tol [2007],* introduces two new modifications of the h-index, in an effort to remove some of its disadvantages. The two modifications are similar to the g-index, however they are based on harmonic and geometric averages instead of arithmetic averages. The f-index is calculated by solving with respect to f the inequality: $\max_f \dfrac{1}{\frac{1}{f}\sum_{i=1}^{f}\frac{1}{c_i}} \geq f$, where $c_i$ denotes the number of citations received by article i (i=1,2,…,n). Similarly, the t-index is calculated by solving the following expression: $\max_t \exp\left[\dfrac{1}{t}\sum_{i=1}^{t}\ln(c_i)\right] \geq t$.

According to the author, it is always: h≤f≤t≤g, and the f- and t-indices have more discriminatory power in comparison to h- and g-indices. A real-data application of these four indices revealed the existence of strong correlations between them, however utilization of the new indices did not change significantly the ranking of the researchers.

*An illustrative example*

To check the relative performance and to make comparisons between the aforementioned h-type indices intended to take into account the differences of research output included in the h-core, we have constructed an artificial example of 7 research outputs (see Table 1).

**Table 1:** Artificial h-core citation oututs of 7 scientists of the same h-index

| j | h-core research ouput of scientists | | | | | | |
|---|---|---|---|---|---|---|---|
| | A | B | C | D | E | F | G |
| 1 | 35 | 150 | 100 | 200 | 85 | 50 | 30 |
| 2 | 34 | 20 | 20 | 20 | 85 | 50 | 25 |
| 3 | 33 | 20 | 20 | 20 | 23 | 50 | 24 |
| 4 | 32 | 19 | 19 | 19 | 20 | 30 | 22 |
| 5 | 31 | 17 | 17 | 17 | 18 | 28 | 17 |
| 6 | 30 | 16 | 16 | 16 | 15 | 25 | 16 |
| 7 | 29 | 14 | 14 | 14 | 14 | 20 | 16 |
| 8 | 28 | 14 | 14 | 14 | 10 | 16 | 14 |
| 9 | 28 | 10 | 10 | 10 | 10 | 11 | 11 |
| 10 | 10 | 10 | 10 | 10 | 10 | 10 | 10 |

The seven scientists – all sharing the same h-index - have varying h-cores. For instance, scientist A is a representative case of a constantly-productive scientist who steadily publishes papers receiving a significant number of citations. On the other hand, scientists B, C and D have one highly-cited publication, that received 150, 100 and 200 citations respectively, raising significantly their h-core citation numbers. Scientists E and F share a pattern that resembles more the pattern of scientist A, with a considerable number of citations concentrated though in the top 2 or 3 highly cited papers. Finally, scientist G is the less-cited scientist among the 7 researchers, with 185 h-core citations.

The following table (Table 2) presents the calculated indices values for the above ficticious example of research output.

**Table 2:** Indices values for the 7 fictius research outputs

|  | indices values of scientists | | | | | | |
|---|---|---|---|---|---|---|---|
|  | A | B | C | D | E | F | G |
| h-index | 10 | 10 | 10 | 10 | 10 | 10 | 10 |
| g-index | 17 | 17 | 15 | 18 | 17 | 17 | 13 |
| A-index | 29 | 29 | 24 | 34 | 29 | 29 | 18,5 |
| R-index | 17 | 17 | 15,5 | 18,4 | 17 | 17 | 13,6 |
| $h_w$-index | 16,7 | 13,8 | 12,6 | 14,1 | 13,9 | 15,3 | 12,2 |
| w-index | 3 | 2 | 2 | 2 | 2 | 3 | 2 |
| i×ci-index | 252 | 150 | 112 | 200 | 170 | 150 | 112 |
| A(2)-index | 5 | 4 | 4 | 4 | 4 | 5 | 4 |
| h-core citations | 290 | 290 | 240 | 340 | 290 | 290 | 185 |

One observes that in the above table the lowest values of all indicators are assigned to scientist G. This is expected since G has the lowest number of citations (185), and these are not equally assigned to the h-core publications but are concentrated on a few top publications. Higher values of the three indices (namely the g, A and R-indices) are assigned to research output of scientist D. Indeed, scientist D has a g-index value of 18, an A-index of 34 and an R-index of 18,4. It has already been reported in previous studies [see, e.g., *Jin et al., 2007; Rousseau, 2006*], that the aforementioned indices are very sensitive to h-core articles receiving an extremely high number of citations. Thus, it is natural in our example for scientist D to be favored by these indices since he has an extremely highly cited paper with 200 citations.

On the other hand, the citation-weighted h-index ($h_w$) of *Egghe and Rousseau [2007]*, manages to better differentiate between the 7 outputs (by ranking first scientist A, and second scientist F). However, the $h_w$ values vary moderately, making the $h_w$-index one of the less variable indices. For instance, the values of the $h_w$-index for scientists C and G are 12.6 and 12.2, respectively. When it comes to the $A^{(2)}$ and w-indices, the calculation of which is based essentially on a similar scheme, it is easily seen that they lack discrimination power, when compared to the previous indices (i.e. the h-, g-, A-, R- and $h_w$-indices). This however is expected, since the two h-type modifications are mainly suitable for large citation outputs and their use is intended to identify and discriminate scientists of significant achievements, from researchers of more common scientific activity. Finally, the $i \times c_i$ index, while manages to rank first scientist A, fails to clearly discriminate the remaining outputs. For instance, it assigns the same index value to scientists C and G ($i \times c_i = 112$), who are clearly different given the presence of an extremely highly cited paper in the output of scientist C.

*4.2. h-type indices for correcting for the age of publications*

*Contemporary h-index, trend h-index, normalized h-index*

*Sidiropoulos et al. [2006]* demonstrate some of the disadvantages of the h-index, and propose a series of generalizations (modifications) of the specific index. They introduce two generalizations of the h-index, the contemporary h-index and the trend h-index, which are modify the h-index in order to reveal the significant young scientists and trendsetters, respectively.

In addition, they define a normalized h-index, which "corrects" for the number of publications, i.e. it gives advantage to the scientists with few but good (largely cited) publications. In the following, each of the proposed indices is presented.

As we mentioned already, the original h-index does not take into account the "age" of the article. It may be the case that a scientist has published a number of significant articles that result in a large h-index, but now he/she is rather inactive or retired. On the other hand, another scientist may still producing significant work. To detect these differences in "time", *Sidiropoulos et al. [2006]* define the contemporary h-index $h^c$ as follows:

A researcher has contemporary h-index $h^c$, if $h^c$ of its $N_p$ articles get a score of $S^c(i) \geq h^c$ each, and the rest ($N_p - h^c$) articles get a score of $S^c(i) \leq h^c$, where $S^c(i)$ is defined as: $S^c(i) = \gamma \times (Y(now) - Y(i) + 1)^{-\delta} \times |C(i)|$, with Y(i) denoting the publication year of article i, and C(i) the articles citing the i-th article. If $\delta=1$, $S^c(i)$ is the number of citations the article i has received divided by the "age" of the article. The coefficient γ is used to "correct" the small value of the derived index, and is suggested to take the value of 4.

For the trend h-index, *Sidiropoulos et al. [2006]* say that a researcher has trend h-index $h^t$, if $h^t$ of its $N_p$ articles get a score of $S^t(i) \geq h^t$ each, and the rest ($N_p - h^t$) articles get a score of $S^t(i) \leq h^t$, where $S^t(i)$ now is given by:

$$S^t(i) = \gamma \times \sum_{\forall x \in C(i)} (Y(now) - Y(x) + 1)^{-\delta}.$$

Apparently, for $\gamma=\delta=1$ the trend h-index coincides with the original h-index. The trend h-index, does not assign a decaying weight to the articles of the researcher, but assign to each citation of the article an exponentially decaying weight. As the authors claim, by doing this the impact of a researcher's work at a particular time instance is measured.

Finally, a researcher has normalized h-index $h^n = h/N_p$, if h of its $N_p$ articles have received at least h citations each, and the rest ($N_p$-h) articles received no more than h citations.

By using real data examples collected from the DBLP database (A server providing bibliographic information on major computer science journals and proceedings, http://kdl.cs.umass.edu/data/dblp/dblp-info.html), the authors calculate the h-index, as well as the previously defined generalizations of it and compare the results.

*The AR-index*

While the R-index of *Jin et al. [2007]* measures the h-core's citation intensity, the AR-index goes one step further and takes the age of each publication into account. This allows for an index that can actually increase and decrease over time. The authors propose the combination of h- and AR-index as a suitable indicator for research work evaluation.

The AR-index is an age-dependent index, built in order to overcome the problem that the h-index always increases even in the case where a scientist stops to produce new work (by simply increasing his/hers citations).

To define the AR-index let $α_j$ denote the age of the article j. Then, the AR-index is defined as: $AR = \sqrt{\sum_{j=1}^{h} \frac{citation_j}{\alpha_j}}$.

The advantage of the AR-index is that it includes in its calculation the age of the articles, thus decreasing when articles become old. In this way the h-index is complemented by an index that can actually decrease. The AR-index is based on the h-index as it makes use of the h-core. *Jin et al. [2007]* present some real examples involving calculations of the R- and the AR-index, in order to show that the two proposed modification indices improve the specific disadvantages of the h-index (and the A-index).

In a more recent work, *Egghe and Rousseau [2007]* present (in both discrete and continuous settings) Jin's indices, by defining a general continuous model. For instance, the A-index of *Jin [2006]* in the continuous setting can be expressed as: $A = \frac{1}{h}\int_{0}^{h} \gamma(r)dr$, where γ(r) denotes the continuous rank-frequency function: γ:[0,T]→[1,+∞]:r→ γ(r).

*The m quotient (or m parameter)*

Initially defined by *Hirsch [2005]*, the m quotient [see also *Bornmann et al., 2008; Imperial and Rodriguez-Navarro, 2007*] is defined as: $m = \frac{h}{y}$, where y denotes the number of years passed since the initial publication of the scientist. Accordingly to *Hirsch [2005]*, a value of m≈1 characterizes a successful scientist, whereas an m-value of approximately 2 and 3 is indicative of an outstanding and a truly unique scientist, respectively. From its definition, it is evident that the m quotient is a useful tool when one needs to compare scientists with different lengths of scientific career.

*4.3. Correcting the h-index for different fields of research*

As stressed out already in section 3, an important disadvantage of the h-index is that typically it cannot take into account the specific field of research of a researcher. In other words, trying to compare the h-indices of two scientists of different fields is not at all a straightforward procedure, since publication rates as well as citation rates vary significantly from one field to another. As reported by *Adler et al. [2008]* (see also *Amin and Mabe [2000]*), the average citations per article in life sciences is about 6 times higher than those in mathematics and computer sciences, making direct comparisons of citation outputs between scientists of these two disciplines invalid. In general, normalization of bibliometric indicators to account for interdisciplinary differences has already been considered in the literature [see, e.g., *van Raan, 2005; Podlubny, 2005; Podlubny and Kassayova, 2006*]. However, relatively little work has been done in this direction, in relation to the h-index and its modifications.

According to Hirsch, scientists in life sciences tend to achieve much higher h-values when compared to scientists in physics. For instance, in physics, a moderately productive scientist usually has an h equal to the number of years of service while biomedical scientists tend to have higher h values [*Hirsch, 2005*].

Thus, prior to comparisons of the h-index, in such situations some kind of "normalization" of the h-indices is required. In this direction, *Iglesias and Pecharromán [2007a]* [see also *Iglesias and Pecharromán, 2007b*] propose a scaling of the h-index to account for the different scientific fields of researchers, assuming a stochastic model for the number of citations (specifically the distribution of the number of citations is assumed to follow Zipf's law), which leads to the following expression for the theoretical h-index: $h = \sqrt[3]{\frac{N_p}{4} \chi^{2/3}}$, where $N_p$ denotes the total number of papers published and $\chi$ is the average number of citations per paper for the researcher. Based on the above specifications, *Iglesias and Pecharromán [2007]* suggest using as a normalizing factor for the h-index the following expression: $f_i = (\chi_{physics} / \chi_i)^{2/3}$, where $\chi_i$ is the average number of citations per paper of scientific field i, and $\chi_{physics}$ (which is the average number of citations per paper for the Physics field) stands as the reference category. Thus, the normalized h-index is given by: $h_{normalized} = f_i \times h = (\chi_{physics} / \chi_i)^{2/3} \times h$.

This normalization methodology is applied to a real dataset consisting of h-index values of highly cited researchers (HCRs) affiliated with Spanish Institutions. The results show that, after correction with the normalizing factor, the h values become more homogeneous. The authors also note that this correction is found particularly useful in the field of mathematics, where HCRs share h-index values considerably lower when compared to HCRs of other disciplines.

*4.4. h-type indices for Journals*

The process of journal evaluation goes back many years in time, and various tools for ranking and comparing journals have been proposed. Nowadays, it is common practice to use the well-established impact factors (IF) as the standard measure of journal quality [*Garfield, 1955; Garfield, 2006*]. The Impact Factor - devised by the Institute of Scientific Information (ISI) – is essentially the average number of citations received within a specific year by articles published in the specific journal in a previous given period of time. Usually, the impact factor of a journal is calculated on information collected within a three-year period. For instance, the IF of a journal for the year 2000 is given by: $IF = \frac{C_{2000}^{1998-99}}{N_{1998-99}}$, where $N_{1998-99}$ is the number of articles published in the specific journal, while $C_{2000}^{1998-99}$ denotes the number of citations of these articles received in 2000.

Recently it has been suggested [see *Rousseau, 2007a; Braun et al., 2005, 2006; Charpon and Husté, 2006,* among others] that the h-index could be used as an alternatively for the ranking of journals. There is a considerable amount of research being carried out on ranking journals according to their h-index, and in the sequel we present some of it.

*The h-index of a journal*

*Braun et al. [2006]* suggest that use of h-type indices in journal ranking could be employed as a supplementary indicator to impact factors because of two important properties of the h-index: its robustness to accidental citations and that it combines quantity (articles published) with impact (citations received). They illustrate – using the Web of Science (WoS) – an easy way of determining the h-index for journals. By using WoS they calculate the h-indices for 21 journals, most of them from the biomedical field (including Nature and Science), and compare these results with the corresponding impact factors of the journals. The results show that the two rankings differ significantly, stressing the different dimensions indicated by the two indices. Further to the work of *Braun et al. [2006]*, *Schubert and Glanzel [2007]* apply the Paretian theoretical model of *Glänzel [2006]* to Braun et al.'s journal citation data.

Other contributions to the subject are due to *Valcnay [2007]* who is also supportive of utilizing h-indices instead of impact factors in journal ratings, given the several "good" properties of the former, such as robustness against possible errors attributed to publications and citations in the tails of the associated distributions, "grey literature" or accidentally counted "highly cited" articles. According to *Valcnay [2007]*, the h-index exhibits two further advantages compared to IF: It is integer-valued, thus avoiding false impresion of precision conveyed by the three decimal points in the IF, and it is much easier to be verified given its simplicity. All the above-mentioned arguments are illustrated by a practical example.

*Rousseau [2007a]* calculates and studies the h-index of the Journal of the American Society of Information Science (JASIS) for the time period between 1991 and 2000. The author observes that the yearly h-index of JASIS is influenced by the number of articles published in the current year, thus he suggests dividing the h-index by the latter number, calling the derived index the relative h-index [see also *Orbay et al., 2007* for a similar application of the relative h-index on data collected on the Turkish Journal of Chemistry].

In another study, *Saad [2006]* examines possible associations between standard indicators of journal impact (i.e. IF) and the h-index for journals. In particular, two datasets including journals from bussiness and marketing were selected to examine correlation coefficients between IF and h for the two sets. The results showed significant correlations between IF and h.

*Miller [2006]* examines correlations between impact factors and h-index values for some of the most popular journals in the field of physics. By observing that the two measures rarely correlate to each other, he deduces that the IF is not an adequate measure of research quality. However, this conclusion is based on the simplified hypothesis that the h-index is unquestionably the global measure of scientific quality, a hypothesis that is questionable [see, e.g., *Adler et al., 2008*].

*Barendse [2007]* investigates performance of journals covering different disciplines by comparing impact factors and a specific modification of the h-index that accounts for the size- and discipline-variability between the journals. The author, like *Rousseau [2007a]*, notices the influence of the journal size to the calculation of the h-index, and proposes a "normalizing" factor of his own, which he calls the strike rate index (SRI), $SRI = \frac{10 \times \log(h)}{\log(N)}$, where N is the total number of articles published by the journal in a given time period. Among other results, the author observes a significant linear relationship between the SR index and the amount of work N published by the journal. The analysis showed that values of SRI rarely correlate with IF values, a result attributed to the general behavior of the two indices. (The h-type index generally favors journals receiving a lot of citations in a long-term period of time (e.g. a 20-year period), while the IF favors journals with articles that receive citations in the first two or three years after publication).

*The impact index $h_m$ for journals/Institutions*

Another more recent application of the h-index in journal ranking can be found in *Molinari and Molinari [2008]*. The authors utilizing data on numbers of papers from three well-known journals from the WoS (Science, Acta Materialia and the Journal of the Mechanics and Physics of Solids), calculate h-index values for the three journals, for various countries. By plotting the derived h-values against the corresponding number of papers for the three journals and the various countries they observed that the plotted points are scattered around a straight line, which they name the master curve (or the m-curve) of the journal considered. Among their empirical findings is that all m-curves considered in the study can be essentially decomposed into two sections, comprised of an initial straight line (corresponding to a relatively small number of published articles), and a second curve (corresponding to a relative large number of published articles). For the large numbers of papers *Molinari and Molinari [2008]* reported that the h-index is associated with the number of published papers, since it can be expressed as: $h = h_m N^{\beta}$, where N denotes the number of papers and β is approximately 0.4 in all cases examined. $h_m = h/N^{\beta}$ corresponds to the point of the m-curve for the specific country of the selected journal, and is called the impact index. The authors, by examining the robustness (especially for large datasets) of the impact index, propose its use for comparing journals. Similar results have been found when implementing the impact index for ranking of Institutions.

Following the work of *Molinari and Molinari [2008]*, *Kinney [2007]* compares the scientific performance of a large number of US Institutions and science centers from the fields of physics and engineering, using data also obtained from the WoS database. The results reveal that the higher $h_m$-values were assigned to the top-ranked US academic Institutions.

*4.5 h-type indices correcting for co-authorship*

As already discussed in the previous section, co-authorship could have a significant impact not only on the value of the h-index but on other bibliometric indicators as well. For instance, *Persson et al. [2008]* found significant correlations between the number of co-authors of a scientist and the mean number of citations per year.

To overcome situations of this nature, *Batista et al. [2005]* divides h by the mean number of researchers in the first h publications (i.e. by the mean number of authors in the papers of the h-core), say $<N_a> = N_a^{(T)}/h$, where $N_a^{(T)}$ is the total numbers of authors in the considered h papers, and called the derived variant of h-index the $h_I$-index. Taking into account that the use of the mean could lead to unfairness towards scientists with a few but largely co-authored articles, *Batista et al. [2005]* propose to divide the h-index by the median number of researchers. The idea of correcting the h-index for co-authorship had been already suggested by *Hirsch [2005]* who proposed the normalization of h-index by a factor that reflects the average number of co-authors. Similar ideas can be also found in *Wan et al. [2007]*, where the actual number of co-authors in a published work is taken into account in calculating a researcher's h-index. The proposed index (called the pure h-index) $h_p$ is obtained by dividing the researchers' h-index with the h-core average number of co-authors, i.e.:

$$h_p = \frac{h}{\sqrt{E(author)}}, \quad \text{where} \quad E(author) = \sum N_E(author, D)/h, \quad \text{and}$$

$N_E(author, D) = 1/S(author_D)$. S is the normalized score of the scientist in paper D [for more details see *Wan et al., 2007*]. Further extensions and improvements of the pure h-index can be found in *Chai et al. [2008]*. The so-called Adapted pure h-index intends to be a less-biased variant of the pure h-index, with respect to authors with many multi-authored papers. The new index, in contrast to the pure index, does not use the h-core for its calculation, but is based instead on a larger number of articles, adapted each time according to the observed citation data.

In another attempt to construct an index that can adequately adjust for the number of co-authors of a scientist in measuring his/hers citation impact, *Schreiber [2008a; 2008b]* devised the $h_m$-index (with subscript m accounting for the multiple authorship of the scientist), based upon the fractionalized counting of the scientist's articles [for details on the fractionalized counting we refer the interested reader to *Egghe et al.,*

*2000*]. In particular, $h_m$ is calculated utilizing a different ranking of articles, which *Schreiber [2008a; 2008b]* calls effective ranking ($r_{eff}$) and is based on the number of co-authors in each article. These ranks are calculated by the following scheme:

$$\left\{ \begin{array}{ll} r_{eff}(1) = 1/\alpha(1), & for \quad r = 1 \\ r_{eff}(r) = r_{eff}(r-1) + 1/\alpha(r), & for \quad r > 1 \end{array} \right\}$$

where α(i) denotes the number of authors of paper i. The $h_m$-value is then calculated in a way similar to the one used to calculate the h-index. To calculate the h-index, we place the number of citations in decreasing order, and then we calculate the h-index by comparing the two columns (articles and citations received). The h-index is the value that corresponds to the article in the h-th position that receives h or more citations. Similarly, the calculation of the $h_m$-index is based on the comparison of the column of citations, and the column of effective ranks (instead of the column of articles). The $h_m$-index is the value corresponding to the number of citations for which the effective rank is equal to or larger than.

Using three fictitious examples and an empirical case of citation output, *Schreiber [2008a]* compares the relative performance of the $h_m$-index to *Batista's et al. [2005]* $h_I$-index, and argues in favor of the $h_m$-index since in contrast to $h_I$, it is more robust to extreme cases of large numbers of co-authors and additionally does not decreases when the number of citations increases. Similar fractional counting approaches based on devising h-type indices for accounting co-authorship can be found in *Egghe [2008c]*.

*4.6 The self-citation issue*

Hirsch index, a tool mainly proposed for the assessment of impact of researchers in the scientific community, in principle should not include self-citations [see, e.g. *Schreiber, 2007b*]. Under this perspective, *Schreiber [2007a]* examines the influence of self-citations on the h-index and distinguishes two kinds of self-citations: the researcher's own citations and the citations made by possible co-authors of the researcher. He argues that while the impact of self-citations is usually insignificant in the h-values of researchers having reached a maturity stage in their careers, it is not negligible in the case of young researchers with "small" citation outputs. To avoid significant distortions of the h-index, *Schreiber [2007α]* suggests excluding at least self-citations of the first type (i.e. citations made by the researcher to his/her own work), especially in situations where the h-index is utilized in Academic evaluation processes (such as academic promotions or new academic positions). *Schreiber [2007b]* examines the influence of self-citations on the g-index proposed by *Egghe [2006b]* and proposes improving it by excluding the self-citations. To verify that self-citations influence the h- and the g-index the author presents nine practical cases in physics where he compares the g- and h-values with and without self-citations. The author argues that the g-index characterizes the dataset better compared to the h-index, and that the influence of self citations is more apparent in the g-index than in the h-index.

Applications of the h-index in the context of self-citations are also given in *Cronin and Meho [2006]* who apply the h index to information science. They calculate the h-index with self-citations included and excluded. Comparison of the two rankings reveals that in general, elimination of self-citations does not much influence the rank ordering of the scientists. As the citation data mainly refered to scientists of mature academic age sharing a large number of citations each, the above findings verify Schreiber's point of view of no major influence of self-citations on the h-indices of influencial scientists.

*4.7 The successive h-index*

Another modification of the h-index and/or its applications across different fields, is the notion of the successive h-index, originally devised by *Schubert [2007]*[6]. The proposed methodology essentially incorporates a hierarchical-type structure in the derivation of h-indices.

This simple idea is based on the calculation of an h-index from the arrangement of a set of other previously calculated h-indices. For instance, the h-index of a University/Institution can be calculated by the following two-step procedure:
- At stage 1 calculate the individual h-indices of the scientific faculty of University/Institution
- At stage 2 arrange the calculated individual h-indices in decreasing order, and apply the definition of h-index to this series, to obtain the successive h-index

---

[6] Independently proposed also by Prathnap [2006].

In this way, an h-index indicative of the overall research output performance of the University/Institution of interest can be obtained. Subsequently to the paper of Schubert, *Egghe [2008a]* studies the successive h-index from a theoretical perspective, by assuming the Lotkaian system of *Egghe and Rousseau [2006]* for modeling the publication/citation distribution, and shows that in each consecutive step of the calculation of the successive h-index, multiplication of the exponent of the previous successive h-index by 1/α is involved, where α denotes a Lotka exponent.

*Egghe and Rao [2008]* take the theoretical model for successive h-indices of *Egghe [2008a]* one step further, by studying it in comparison to two other indices, the $h_p$ and $h_c$, proposed by the authors. Indices $h_p$ and $h_c$ correspond to the h-index that is calculated by arranging in decreasing order publications and citations of individuals, respectively. By utilizing the Lotkaian model they show that the following inequality for the three indices holds: *successive* $h < h_p < h_c$. An application on a sample of 167 researchers from the field of optical flow estimation concludes the study.

Another application of the use of successive h-indices can be found in *Arencibia-Jorge et al. [2008]*. They are used for the assessment of scientific performance of the Cuban National Scientific Research Center (for the period 2001-2005) as well as of the scientific performance of the Departments of the Institute.

*4.8 h-index sequence and h-index matrix*

An attempt to overcome the inefficiency of the h-index in not taking into account basic aspects of a researchers publication output, such as the scientific age of a researcher or the stability of the quality work throughout his career, was made by *Liang [2006]*. The author proposes two alternatives for presenting h-index values of a researcher, namely the h-index sequence and the h-index matrix, aiming at revealing differences between academic careers of researchers not easily identified by the single number of the h-index. As suggested by its name, an h-index sequence is a sequence of h-indices, the first of which is calculated starting from the publications and received citations of the year of the last available publication of the researcher, and continuous with the calculation of the h values for each preceding year. Thus, if say t denotes the year of the last article published by researcher A, the h-sequence will be: $h_t$, $h_{t-1,t}$, $h_{t-2,t}$, ..... $h_{t-k,t}$ where t-k is the year of the first publication of the researcher. Accordingly, the h-matrix is created by arranging all h-sequences of the researchers' of interest, intending to make comparisons of scientists at various levels of their scientific career. *Liang [2006]* using data on 11 physicists, constructs their h-index sequences and compares them to derive useful results associated with the different patterns on these sequences, indicating that progress and trends of the h-index throughout the career of scientists varies significantly from scientist to scientist.

However, despite the novelty of the idea, the author restricts his findings to the observation of the specific eleven h-sequences, without employing any statistical analysis to generalize the specific findings and behavior patterns of the progress of the h-index through time to population by using of an adequate statistical model. In addition to time effects, investigation of interactions between time and a variety of other factors on the progression of the h-index during the course of the career of scientists could help to obtain significant information about the overall profile of the work of researchers over time.

*4.9 The research status index*

In another setting, *Symonds et al. [2006]*, examine the publication records of 168 scientists in the field of ecology and evolutionary biology in order to assess gender differences in research performance. According to the authors, the h-index is strongly biased against female researchers. They propose a modified index to correct for this bias, in order to assess research performance of male and female researchers on a more equal basis. The authors follow the publication record of 39 female and 129 male researchers of the life sciences departments of British and Australian Universities. Using the Web of Science they counted the number of publications and the number of citations of each of the publications.

Consistently with previous studies, the authors observed a clear difference in the number of publications produced by males and females, with men publishing on average almost 40% more papers than women. As concerns the h-index, it was found to favor less the female scientists.

As a remedy, the authors introduce an alternative metric to h-index, namely the residual h, which they call *research status*. It is calculated as the y-residual from the least squares regression line of h on the number of publications. Calculation of the research status for the data set already described showed no difference between male and female researchers.

The authors conclude by presenting some disadvantages of the Research Status index, namely that it is affected by the addition of a small number of low cited papers, and that it appears to completely disregard the quantity of research.

*4.10 The h-b index for topics or compounds*

*Banks [2006]* developed the idea of using an h-type index to "measure" impact of a scientific topic (or compound) and referred to the latter index as the h-b index. The main purpose of the h-b index is to reveal and separate the interesting topics in scientific research ("hot" topics, that young researchers at the start of their career or PhD students would like to know in order to advance with their work) from topics of no scientific interest or topics in which a lot of work has been already done and are now exhausted. The h-b index can be calculated from the Thomson Web of Knowledge by entering in the search field not a scientist's name but a selected scientific topic of interest.

**5. Theoretical Approximations to the h-index**

While a long series of h-type modification indices have been proposed in the literature, and a significant number of practical implementations of the h-index have appeared, the mathematical/statistical properties and behaviour of the index has not been examined in full depth yet. In *Hirsch [2005]*, one may find an early attempt to analyze the properties of the theoretical h-index, through the presentation of a simple deterministic model. Only recently, attempts have been made in this direction [see for instance *Burrell, 2007a; Glänzel, 2006; Egghe and Rousseau, 2006*].

*Glänzel [2006]*, attempts to interpret theoretically some properties of the h-index, having assumed a citation distribution, using extreme value theory. He analyzes the basic properties of the h-index on the basis of a probability distribution model (using the Pareto distribution). *Glänzel [2006]* defines the theoretical h-index (which he denotes by H), using Gumbel's characteristic extreme values *[Gumbel, 1958]*. Under this setting, if X is a random variable, with cumulative distribution function (CDF) F(k)=P(X<k), then Gumbel's rth characteristic extreme value is defined by: $u_r = G^{-1}(r/n) = \max\{k : G(k) \geq r/n\}$, where G(k)=1-F(k), and n is a given sample from a population following distribution F. Then, the (theoretical) H-index is defined as: $H = \max\{r : u_r \geq r\} = \max\{r : \max\{k : G(k) \geq r/n\} \geq r\}$.

The author studies two examples using the discrete Pareto distribution and the Price distribution (a special type of a Paretian distribution).

*Schubert and Glänzel [2007],* test the theoretical model of *Glänzel [2006]* in practical implementations using journal citation data, collected from the Web of Science database. They concluded that the theoretical Paretian model fitted perfectly to the data collected from journals.

*Burrell [2007a],* proposes a simple stochastic model in order to investigate the h-index and its properties. His parametric model distinguishes between an author's publication process and the subsequent citation process of the published papers. The number of publications is assumed to follow a Poisson distribution, while the citation rate (i.e. the mean number of citations per unit time) is taken to follow a gamma distribution (we refer the interested reader to *Burrell [1992]* for more details on this stochastic publication-citation model). Under the latter assumptions, the author provides the distribution of the number of citations as well as the formula for the expected value of the distribution. By exploring different scenarios using various values for the model's parameters, he applies the theoretical model to simulated data, and finds that the (theoretical) h-index is approximately proportional to the author's career length, and approximately linearly related to the logarithms of the author's productivity rate and average citation rate. Finally, an application of the stochastic model of *Burrell [2007a]* is provided, along with an investigation of the associations between the h-index, *Jin's [2006]* A-index and the h-core of the h-index. Using regression analysis, the author showed that the A-index is linearly related to the h-index and time, and the h-core is linearly related to $h^2$ (and consequently linearly related to $A^2$).

In another theoretical context, that of an Information Production Process[7] *[see Egghe, 2005], Egghe and Rousseau [2006],* using a source-item terminology, show that if a system has T sources and a Lotka function exponent α, the system's unique h-index is given by the expression: $h = T^{1/\alpha}$. Moreover, relations between h, T and α are examined in depth.

*Egghe [2008b]* takes the work of *Egghe and Rousseau [2006]* one step further, by incorporating the notion of time in the latter expression of the h-index for information production processes, showing that the time-dependent h-index (refered by the author as the dynamic h-index) can be written as: $h = \left[\left(1-b^t\right)^{a-1} T\right]^{1/a}$, where T is the total number of articles, α is Lotka's exponent, and b denotes the ageing rate of citations. This expression, for t→∞, reduces to the expression of *Egghe and Rousseau [2006]*.

---

[7] In this context, sources are equivalent to articles published, whereas the produced items correspond to the received citations.

As already mentioned, while there exists a vast literature on the empirical h-index and its applications, relatively little work has been done on the study of the theoretical h-index as a statistical function, allowing to construct confidence intervals, test hypotheses and check the validity of its statistical properties. Recently, *Beirlant and Einmehl [2007]* establish the asymptotic normality of the theoretical h-index under a non-parametric framework. Furthermore, the authors apply their general results assuming two well-known distribution functions (Pareto and Weibull distributions) to the number of citations and construct confidence intervals for the empirical h-index. Finally, the proposed methodology is illustrated by two practical examples, using citation data on two distinguished researchers, namely D.R. Cox and P. Erdös.

## 6. Other Developments Related to the h-index

*Bornmann and Daniel [2007],* provide an illustration of the advantages and disadvantages of the h-index. Subsequent corrections and complements to the h-index are also presented.

A study by *Lehmann, Jackson and Lautrup [2005]* argues against the accuracy of the h-index for measuring scientific performance. By presenting a general Bayesian method for quantifying the statistical reliability of some one-dimensional measures such as the h-index, the authors deduce that the h-index is shown to lack the necessary accuracy and precision in order to be useful. On the other hand, the statistical analysis performed showed that the mean, median and maximum numbers of citations are reliable and permit accurate measurement of scientific performance.

J.E. Hirsch, in his latest article on the h-index *[Hirsch, 2007],* discusses the possibility of predicting future work of a researcher using the h-index. In doing this, *Hirsch [2007]* employs data from the ISI Web of Science database, from various time windows of a researcher's publishing life and examines the significance of correlations of h-indices between these time frames. For comparison, some other indicators such as the total number of citations ($N_c$), the total number of publications ($N_p$) and the citations per paper are examined. The results of the analysis show that h and $N_c$ are better in predicting a researcher's future achievements, when compared to $N_p$, whereas the h-index is found to be slightly superior in comparison to $N_c$. The author argues that the superiority of the h-index is mainly attributed to the presence of co-authoship, an issue which we have already discussed in previous sections. Finally, Hirsch defines an improvement of the h-index, in terms of an expresion that best predicts citation output of a researcher in a future time frame, given by the expression: $h_\alpha = \sqrt{h^2 + \alpha N_c}$, where the coefficient α is approximately equal to −0.1. This expression tells us, that between two scientists having the same h-indices but unequal number of citations at the present time, the one expected to have a higher number of citations in the future is the one with the lower number of citations presently (a paradox attributed by Hirsch to co-authorship).

*Vinkler [2007]*, argues against the suitability of using a single measure (such as the h-index) for measuring the productivity of a researcher, and suggests the use of several indicators weighted for the purpose of the assessment. By presenting a series of simple measures (dependent on articles published and citations received) he examines – through a data application – and finds significant correlations between the h-index and one of the proposed indices. He concludes that the h-index is not appropriate for the assessment of research performance of scientists publishing low numbers of articles.

In a comparative study of some of the most important h-type indices proposed in the literature, *Bornmann et al. [2008]* perform an exploratory factor analysis (EFA) using as observed variables nine h-type indicators (including the h-index) in an effort to reveal the latent factors causing the latter indices. The study concludes that more than 95 per cent of the variability in the factor model is explained by two factors. The first factor was recognized to describe the h-core (including the h-, g-, and $h^{(2)}$-indices among others), whereas the second factor was recognized to describe the impact of the papers in the h-core (including the A-, R-, and AR-indices among others). Finally, logistic regression analysis is employed to predict peer assessment, from the two factors, used as independent variables in the logistic model.

*Costas and Bordons [2007]*, implement exploratory factor analysis to investigate possible associations of the h-index with other measures of scientific research (measures that describe both quality and quantity of the performance of a researcher), using data on the publication/citation output of Spanish scientists in the field of natural resources (the data cover the period between 1994 to 2004 and are available through the WoS). The four factors extracted from the factor analysis explained 93% of the total variance in the data. The point of interest of the analysis is that the first factor (explaining 29% of total variability) comprises the h-index, the number of publications and the number of citations received, while the remaining three factors consisted of relative indicators of quality and quantity. Since the number of citations and publications are characterized as absolute indicators of quantity and impact, respectively, according to the authors' opinion the h-index is confined to explain only a small portion of the information about a researcher's work, leaving unexplained other important aspects of scientific performance, conveyed by the other relative indicators.

Confirmatory factor analysis (CFA) could be also utilized to reveal important aspects of scientists' research outputs. Specifically, by utilizing the comparative advantage of CFA over EFA of allowing for testing hypotheses about a particular model structure, one could impose a hypothesized model of observed variables such as the h-index, Hirsch-type indices and other common bibliometric indicators to derive overall unique measures of scientific research performance, comprising thus properties of each of these measures in one or more multidimensional latent factors.

Another practical application associated with the h-index can be found in *Torro-Alves et al. [2007]*, where the index is utilized for the bibliometric evaluation of the Departments of a Brazilian University and associated programs offered by the departments (both undergraduate and graduate). The results showed that an evaluation based on the h-index performs better when comparing graduate research programs than undergraduate research programs. The findings also reveal the inappropriateness of using a non-normalized index in comparing scientific performance of scientists of different disciplines and the insufficiency of scientific databases to cover adequately citation and publication outputs of researchers of specific fields (e.g., psychology).

*Rousseau [2007b],* examines the influence of missing publications in the calculation of the h-index. Using a theoretical model and assuming that the citations follow the Zipf distribution, *Rousseau [2007]* found that the h-index remains generally unaffected to small numbers of missing "highly-cited" articles, especially when compared to the influence of missing publications.

*Van Raan [2006],* studies correlations between the h-index and several other standard indicators (such as the number of publications, the number of citations excluding self-citations, etc) as well as peer judgments, based on data of an evaluation study conducted on 147 university chemistry research groups in the Netherlands during the period 1991-2000. Among other empirical findings using regression analysis van Raan found that the number of citations is proportional to the square of the h-index [see *Hirsch, 2005; van Raan, 2006*], and that $N_{citations}=\alpha h^2$, where α is a constant. Specifically, the author found the relation $h= 0.42 \times N_{citations}^{0.45}$ using data from chemistry scientists. (According to Hirsch the constant α for the discipline of Physics ranges between 3 and 5).

Finally, *Kelly and Jennions [2006]*, examine several factors that might influence the h-index, such as gender, age, country of residence, discipline (or even sub-discipline effects) and the total publication output. By fitting a regression model, the correlation between h and the scientific age normally expected to exist, is verified. Also, controlling for scientific age, it was found that females have lower h-index compared to male scientists. This might be due to discrimination against cited papers by female authors, or indicating that females publish papers that are less citable, or that females publish fewer papers in general. The authors also examine the influence of self-citations.

## 7. A New Classification for Index Comparisons

All the above described indices are intended to "estimate" (better measure) a scientist's impact as concerns his scientific research, mainly based on measuring two basic characteristics: a) the number of articles published and b) the number of citations received by these articles. According to *Cole and Cole [1973]*, researchers can be categorized in the following four basic categories: those who publish many papers and receive many citations, those who publish many papers and receive a small number of citations, those who publish a small number of papers and receive many citations and those who publish a small number of papers and receive a small number of citations.

In addition to these two classifications (i.e., small/significant number of publications, small/significant number of citations), we introduce another feature that we believe is of interest, when assessing the scientific performance of researchers, which is the spread of a scientist's work. For example, a scientist who has published a large number of articles, with a significant number of citations that are not all concentrated on a few of his articles should have a large index value, while a scientist that has published a small number of articles with a few citations, concentrated on only a few of his articles ("highly-cited" articles) should have a small index value. To better demonstrate how the combination of these three basic characteristics influences the values of the indices we present an artificial example of research work (see Table 3). We have constructed eight lists of publications corresponding to eight authors based on combinations of the three characteristics. For instance, the author who corresponds to combination **ACE** is assumed to have published a significant number of articles (say 20), followed by a large number of citations (200) which correspond to only a few, highly cited papers (articles 1, 2 and 3). The author corresponding to combination **BDE** has published a smaller number of papers (10), which are not highly cited (50 citations) and also correspond to a few highly cited papers. It would be natural, by comparing the two authors using a bibliographic index, to rank the first author higher than the second.

**Table 3:** An example of the publication record of 8 researchers according to the three-factor classification scheme

| articles | ACE | ACF | ADE | ADF | BCE | BCF | BDE | BDF |
|---|---|---|---|---|---|---|---|---|
| 1 | 100 | 17 | 10 | 4 | 60 | 25 | 20 | 8 |
| 2 | 28 | 17 | 8 | 4 | 60 | 25 | 12 | 7 |
| 3 | 25 | 17 | 8 | 3 | 50 | 25 | 10 | 7 |
| 4 | 7 | 15 | 7 | 3 | 20 | 20 | 4 | 5 |
| 5 | 7 | 12 | 6 | 3 | 5 | 20 | 2 | 5 |
| 6 | 7 | 11 | 5 | 3 | 1 | 19 | 2 | 5 |
| 7 | 7 | 10 | 1 | 3 | 1 | 18 | 0 | 5 |
| 8 | 4 | 10 | 1 | 3 | 1 | 16 | 0 | 5 |
| 9 | 4 | 10 | 1 | 3 | 1 | 16 | 0 | 2 |
| 10 | 1 | 10 | 1 | 3 | 1 | 16 | 0 | 1 |
| 11 | 1 | 9 | 1 | 3 | | | | |
| 12 | 1 | 9 | 1 | 2 | | | | |
| 13 | 1 | 9 | 0 | 2 | | | | |
| 14 | 1 | 9 | 0 | 2 | | | | |
| 15 | 1 | 9 | 0 | 2 | | | | |
| 16 | 1 | 6 | 0 | 2 | | | | |
| 17 | 1 | 6 | 0 | 2 | | | | |
| 18 | 1 | 6 | 0 | 1 | | | | |
| 19 | 1 | 4 | 0 | 1 | | | | |
| 20 | 1 | 4 | 0 | 1 | | | | |
| | 200 | 200 | 50 | 50 | 200 | 200 | 50 | 50 |
| | Many articles | Many articles | Many articles | Many articles | Few articles | Few articles | Few articles | Few articles |
| | Many citations | Many citations | Few citations | Few citations | Many citations | Many citations | Few citations | Few citations |
| | Closely distributed citations | Widely distributed citations | Closely distributed citations | Widely distributed citations | Closely distributed citations | Widely distributed citations | Closely distributed citations | Widely distributed citations |

Based on the above artificial data we calculated four of the indices already proposed in the literature, namely the h-, g-, A- and R-indices.

**Table 4:** h-index and h-type indices for the publication/citation outputs of the 8 authors

| AUTHOR | h-index | g-index | A-index | R-index |
|---|---|---|---|---|
| ACE | 7 | 13 | 25.9 | 13.5 |
| ACF | 10 | 12 | 12.9 | 11.4 |
| ADE | 5 | 6 | 7.8 | 6.2 |
| ADF | 3 | 3 | 3.7 | 3.3 |
| BCE | 5 | 10 | 39 | 14 |
| BCF | 10 | 10 | 20 | 14.1 |
| BDE | 4 | 7 | 11.5 | 6.8 |
| BDF | 5 | 6 | 6.4 | 5.7 |

As we observe from the above table, ranking of the scientists based on the various proposed indices is not the same for all indices. We see that the h-index places at the top scientists **ACF** and **BCF**, that is the scientists with many and widely distributed citations, and this is not influenced by the amount of work (20 and 10 articles respectively). The h-index gives a low ranking to authors with few citations, and is independent of the amount of work of the authors. For instance, **BDE** has h-index equal to 4, while **ADF** has the lower h-index with value 3, lower than that of **BDE**. However, author **ADF** has twice the number of articles of **BDE** and his citations are more scattered.

With regard to the g-index, we see that it favours authors that produce a large number of articles, with many corresponding citations (**ACE, ACF**).

The A-index, which takes advantage of the h-core for its calculation, favours scientists **BCE** and **ACE**, who have many citations that are concentrated at the h-core. This is an obvious disadvantage of the A-index which, because of its construction, punishes scientists with a higher h-index (since the A-index requires in its calculation the division of the h-core citations by h). For example, among two scientists with the same number of citations in their h-core, say $N_{core\ citations}$, the A-index will assign a higher value to the scientist with the smaller h-index since $A = \dfrac{N_{core\ citations}}{h}$.

Finally, the R-index places at the top authors **BCE** and **BCF**, i.e. authors with fewer publications but with many citations. The specific index was proposed as an alternative to the A-index.

**8. Some Modifications**

In this section, we propose two new modifications of existing indices for measuring scientific impact and we calculate them using the previous artificial data set. The first index is a modification of the R-index, and is given by the following formula: $R_m = \sqrt{\sum_{j=1}^{h} citation_j^{1/2}}$.

That is, $R_m$ is the square root of the summation of the square roots of the citations belonging to the h core.

The R-index has been proposed in order to eliminate the problem of the A-index punishing among scientists with high h-index the ones with more scattered citations (i.e. with greater h-core). However, there are situations where the R-index does not favour the scientist with the greater h core, for instance see scientists **ACE** and **ACF** of our example. The value of the R-index for scientist **ACE** is 13.5, higher of the 11.4 value of the index for scientist **ACF**. Thus, it is clear that in the specific example, the R-index fails to correct the disadvantage of the A-index, favouring the scientist with most citations concentrated to only a few articles.

Another manifestation of the above mentioned disadvantage is seen in the comparison of scientists **BCE** and **BCF**. The R-index is 14.1 for **BCF** and 14 for **BCE**. Here, although the R-index corrects the problem of the A-index and does not give a higher value to scientist **BCE**, it still does not give a significant advantage to the scientist **BCF** who has more scattered citations. This is due to the fact that scientist **BCE** has only a few citations outside his h-core, and thus both h-core sums are approximately equal.

The resulting ranking based on our modification of the R-index is presented in the table below.

**Table 5:** R- and $R_m$-indices for the publication/citation outputs of the 8 scientists

| AUTHOR | $R_m$-index | R-index |
|---|---|---|
| ACE | 5.56 | 13.5 |
| ACF | 5.97 | 11.4 |
| ADE | 3.73 | 6.2 |
| ADF | 2.39 | 3.3 |
| BCE | 5.41 | 14 |
| BCF | 6.67 | 14.1 |
| BDE | 3.62 | 6.8 |
| BDF | 3.55 | 5.7 |

As we observe, the new index $R_m$ slightly improves the ranking of the scientists. It puts at the top scientists **BCF** and **ACF** (i.e. the scientists with a lot and widely scattered citations) and at the bottom scientists **ADF** and **BDF** (i.e. scientists with a few and widely scattered citations). Also, the index values for **ACE** and **ACF** as well as for **BCE** and for **BCF** differ, thus distinguishing between the two pairs of scientists by giving higher values to the scientists with more widely accepted work.

All the indices we studied so far do not take into consideration the variability of the citations. For instance, the A-index is considering the average citations in the h-core, while the R-index is based on the sum of citations included in the h-core. The same holds true for the h- and g-index. However, it is also important to include in the bibliographic measurement this variability, since scientists with less variable citations in their h-core should be rewarded when compared to scientists with a more variable h-core. To adjust for this, we choose to use the well-known coefficient of variation (CV), that has the appealing property taking into account the variability between data of various magnitudes with reference to the central tendency (since by dividing by the mean, the latter is eliminated as a factor).

To adjust for the variability of the h-core citations we formulate a new index by simply subtracting the h-core CV from the index already presented above. The results for our artificial example are presented in Table 6.

**Table 6:** $R_m$- and $R_{m-cv}$-indices for the publication/citation outputs of the 8 researcers

| AUTHOR | h-core cv | $R_{m-cv}$ - index | $R_m$-index | R-index |
|---|---|---|---|---|
| ACE | 1.31 | 4.25 | 5.56 | 13.5 |
| ACF | 0.25 | 5.72 | 5.97 | 11.4 |
| ADE | 0.19 | 3.54 | 3.73 | 6.2 |
| ADF | 0.16 | 2.23 | 2.39 | 3.3 |
| BCE | 0.64 | 4.77 | 5.41 | 14 |
| BCF | 0.19 | 6.48 | 6.67 | 14.1 |
| BDE | 0.57 | 3.05 | 3.62 | 6.8 |
| BDF | 0.21 | 3.34 | 3.55 | 5.7 |

In this way, a penalty has been assigned to the more variable h-core data, making the differences between scientists more visible. For instance, the difference between scientists **ACE** and **ACF** is now greater, as it should be, given that scientist **ACE** has citations concentrated at the top of the h-core (and thus a higher coefficient of variation), while scientist **ACF** has a significantly smaller coefficient of variation since his citations at the h-core are more widely scattered.

## 9. Conclusions

In this article we have presented an extensive and critical review of the existing literature on the h-index and its most important modifications, as well as on other indicators of research output. Furthermore, we have presented some modifications of the h-index, aiming at improving its performance in special circumstances.

Overall, as a general guideline for assessing the citation impact of a researcher, we suggest a combined use of the h-index with other h-type indices for more representative results. In particular, we recommend the use of the h-index along with the $h_w$, g, R and A-index values, to identify significant variations in the h-core outputs. When assessing outputs of senior researchers (or of researchers of significant achievements, such as highly cited researchers), the addition of h-type indices such as the $A^{(2)}$, w and $i \times c_i$-indices can provide extra useful insight. Use of the above measures, when combined with information provided by other standard bibliometric measures (e.g. total numbers of publications/citations) or other criteria of scientific assessment such as peer reviews, can significantly improve the validity of the results provided by the single value of the h-index. In addition, one has to pay attention to the exclusion of self-citations, especially in small-citation sets, in order to improve the accuracy and fairness of the resulting assessment. Effects of co-authorship, mainly when measuring scientific impact in dischiplines such as medicine, where multiple co-authorship is a rather common phenomenon, should also be taken into consideration (for instance, by utilizing the h-type indices presented in this article to correct for co-authorship). Finally, citation data obtained from the various citation sources should be used with caution, regarding the validity of the data provided. Comparisons between citation outputs from different sources may provide significant help in coming up with more credible results.